# A coupled FE-RRM-based numerical model for analysis of energy transmission loss through stiffened double-wall panel due to TBL excitation

Biplab Ranjan Adhikary[a], Atanu Sahu[b], Partha Bhattacharya[a,*]

[a]Jadvapur University, Kolkata, India 700 032

[b]National Institute of Technology Silchar, Silchar, India 788 010

*p_bhatta@daad-alumni.de

A B S T R A C T

We propose a fully coupled numerical model to predict energy transmission through a turbulent boundary layer (TBL) excited stiffened double-leaf flexible aircraft panel using a finite element (FE) framework. Mindlin's first order shear deformation model is adopted for the panels and a TBL-structure-acoustic coupling model is developed using finite element-radiation resistance matrix (FE-RRM) approach to predict the transmission loss (TL) through double-leaf panels with variable thickness and stiffener orientation. The model is also capable to capture the contribution of orthotropic lamina sequence and frequency-dependent structural damping in predicting the TL. Thus, a new numerical model is proposed that enables the designers with greater flexibility in terms of the number of panel leaves, boundary, and stiffening condition of the aircraft panel-cavity-panel system, made of isotropic or orthotropic laminates.

## 1. Introduction

Turbulent boundary layer (TBL) induced vibro-acoustic response for flexible stiffened structural panels as reported so far using turbulence-structure coupling models are estimated in the wave number-frequency domain using the analytical modal expansion technique [1-3]. While FE-based numerical studies, e.g., [4] are confined to single-leaf panel response prediction only, analytical vibro-acoustic models were extended to estimate unstiffened single-leaf [1-2], unstiffened double-leaf [5], and stiffened single-leaf panel-radiated sound power in the free field. However, in real life, the aircraft skin-trim double wall panel configurations have stiffened skin, the results of which are non-existent.

Hence, in this work, a TBL-induced vibro-acoustic response prediction model for the stiffened double-leaf panel is proposed to estimate energy transmission through a panel-cavity-panel system with generic boundary conditions, stiffener position, etc., using a numerical FE-RRM approach, coded in-house in a MATLAB (ver. R2013b) environment.

A zero-pressure gradient TBL is considered in this model and the single-point wall-pressure spectrum is calculated using the Efimtsov 2nd model [6]. Modified Corcos [4] and Mellen [8] models for spatial coherence function are used to measure pressure cross-PSD. Subsequently, panel response is calculated through turbulence-structure coupling. The structural panels are discretized using N number of finite elements. Considering each element behaving as an elemental radiator RRM is developed in the FE framework. Finally, the flow-induced structural response and the RRM are coupled to estimate the radiated sound power (RSP) in the free field. RSP for unstiffened/stiffened single/double panel systems provides a comparative insight into the transmission efficiencies.

**2. Mathematical formulation**

*2.1. Modelling of the turbulent flow field*

The turbulent flow field is considered to be homogeneous and stationary. Single-point TBL wall-pressure spectrum ($\Phi_p$) is calculated using Efimtsov 2nd model because as reported by Thomson and Rocha [9] Efimtsov 2nd model and Reckl and Weston model gives the best prediction of the in-flight wall-pressure PSD in the low-mid frequency (0-800Hz) regime. The Efimtsov 2nd model used is given in Eq. (1).

$$\Phi_p(\omega) = 2\pi\alpha U_\tau^3 \rho^2 \delta \frac{\beta}{(1+8\alpha^3 Sh^2)^{\frac{1}{3}} + \alpha\beta Re_\tau \left(\frac{Sh}{Re_\tau}\right)^{\frac{10}{3}}} \quad (1)$$

where, parameters have their standard meaning and formula, as Eq. (5) to Eq. (10) of Thomson and Rocha [9].

The excited panel is first discretized using a 2D grid and the cross-power spectral density (cross-PSD) over all the grid points is calculated using the coherence function as per the modified Corcos model [7],

$$\Phi_{pp}(x_\mu, x_\nu, \omega) = \sqrt{\Phi_p(x_\mu, \omega)\Phi_p(x_\nu, \omega)} \Gamma(\xi_1, \xi_3, \omega) \quad (2)$$

where, the spatial correlation function is expressed following the combination of the modified Corcos model [4] and Mellen elliptical model [8] as,

$$\Gamma(\xi_x, \xi_z, \omega) = \left(1 + \alpha_x \left|\frac{\omega \xi_x}{U_c}\right|\right) e^{-\alpha_x \left|\frac{\omega \xi_x}{U_c}\right|} e^{i\frac{\omega \xi_x}{U_c}} e^{-\alpha_z \left|\frac{\omega \xi_z}{U_c}\right|} \qquad (3)$$

$$\Gamma(\xi_x, \xi_z, \omega) = \begin{cases} \Gamma(\xi_x, \xi_z, \omega), & if\ \sqrt{\xi_x^2 + \xi_z^2} < \sqrt{L_x^2 + L_z^2} \\ 0, & if\ \sqrt{\xi_x^2 + \xi_z^2} > \sqrt{L_x^2 + L_z^2} \end{cases} \qquad (4)$$

$\xi_x$ and $\xi_z$ are separation vectors between two points, and $L_x = \frac{U_c}{\alpha_x \omega}$ and $L_z = \frac{U_c}{\alpha_x \omega}$ are coherence lengths in the streamwise and cross-stream directions, respectively. Corcos model [7] constants are taken as $\alpha_x = 0.11$, $\alpha_z = 0.70$. [4]

### 2.2. Turbulence-structure coupling

The structural panel(s) is/are discretized using 4-node iso-parametric elements and the discretization is so done that the pressure grid and the structural FE grids coincide and properly mapped.

#### 2.2.1. Single-wall response

For a single panel system, the TBL-induced structural response can directly be estimated as [2],

$$S_{ww}(\omega) = H_w^*(\omega) S_{pp}(\omega) H_w^T(\omega) \qquad (5)$$

Here, $S_{ww}(\omega)$ is the panel displacement PSD, $H_w(\omega)$ is the response function of the panel and $S_{pp}(\omega)$ is the TBL cross PSD of the force, calculated as,

$$S_{pp}(\omega) = A_\mu \Phi_{pp} A_\nu \qquad (6)$$

$A_\mu$ and $A_\nu$ are the elemental areas around the nodes $\mu$ and $\nu$.

#### 2.2.2. Double-wall response; Panel-cavity-panel system

It must be noted that the gap cavity in the double-leaf panel system is modelled using 8-node octahedral FE with pressure DOF. The external force-induced response of the entire structural system can be expressed as,

$$Y(\omega) = H(\omega) X(\omega) \qquad (7)$$

Here, $H(\omega)$ and $X(\omega)$ are the frequency-dependent transfer function and forcing function, respectively.

$$H(\omega) = \begin{bmatrix} H_{11} & H_{12} & H_{13} \\ H_{21} & H_{22} & H_{23} \\ H_{31} & H_{32} & H_{33} \end{bmatrix}^{-1} \qquad (8)$$

Designation: 1-skin panel (panel 'a'); 2-trim panel (panel 'b'); 3-cavity

The uncoupled and coupled transfer functions $(H_{ij})$ are derived from the dynamic equations of the two panels and cavity in modal domain, as detailed in Eq. (5), Eq. (6) and Eq. (12) of Ghosh and Bhattacharya [10].

Subscripts, i = j represent auto coupling, and i ≠ j represent cross-coupling. As there is no direct connection between the two panels, $H_{12} = H_{21} = 0$. The TBL forces act on the skin panel only, and hence the force vector can be written as,

$$X(\omega) = \begin{Bmatrix} F_{tbl} \\ 0 \\ 0 \end{Bmatrix} \quad (9)$$

On assembling Eq. (8) and Eq. (9) as described in Eq. (7) and rearranging, one obtains the response function of the panels ($H_{w,a}$ and $H_{w,b}$) and the cavity ($H_p$) for the coupled system,

$$H_p = (H_{31}H_{11}^{-1}H_{13} + H_{32}H_{22}^{-1}H_{23} - H_{33})^{-1}H_{31}H_{11}^{-1} \quad (10)$$

$$H_{w,a} = H_{11}^{-1}(I - H_{13}H_p) \quad (11)$$

$$H_{w,b} = -H_{22}^{-1}H_{23}H_p \quad (12)$$

The modal response functions are so arranged that they can be solved for any number of modes for panel 'a', panel 'b' and cavity. Once, the response functions are obtained in a coupled system, the panel displacement PSD ($S_{ww,a}$ & $S_{ww,b}$) and cavity pressure PSD ($S_p$) can be calculated as,

$$S_{ww,a}(\omega) = H_{w,a}^*(\omega)S_{pp}(\omega)H_{w,a}^T(\omega) \quad (13)$$

$$S_{ww,b}(\omega) = H_{w,b}^*(\omega)S_{pp}(\omega)H_{w,b}^T(\omega) \quad (14)$$

$$S_p(\omega) = H_p^*(\omega)S_{pp}(\omega)H_p^T(\omega) \quad (15)$$

The panel and cavity responses can be used to estimate cavity pressure, skin panel response, etc. But as the present work is focused on the estimation of the RSP from the trim panel, the displacement PSD, $S_{ww,b}$ is considered and transformed into velocity PSD, $S_{vv,b}$ as given,

$$S_{vv,b} = \omega^2 S_{ww,b} \quad (16)$$

*2.2.3. Stiffened panel*

The stiffeners are modeled using 4-node 2D plate elements and assembled with the panel using necessary transformation, unlike the 1D model as used by Zhou et al. [3].

*2.3. Radiation resistance matrix (RRM) and sound radiation*

In case of harmonic time-dependent acoustic wave propagation through a homogeneous and elastic fluid, the wave equation reduces to Helmhotz differential equation, which for a baffled flat plate, reduces to Rayleigh's second integral in the form,

$$p(\mathbf{r}) = \frac{i\omega\rho_0}{2\pi} \int_S v_n(r_S) \frac{e^{-ik|r-r_S|}}{|r-r_S|} dS \qquad (17)$$

Eq. (17) is solved using a numerical technique by discretizing the entire plate in N number of planar elements (elemental radiator) that are small compared to the acoustic wavelength. The numerical operation leads to RRM, $[\mathbf{R}]$ calculation as described by Ghosh and Bhattacharya [10]. Finally, RSP is calculated as [2],

$$RSP(\omega) = \{\mathbf{v}\}^H [\mathbf{R}]\{\mathbf{v}\} \qquad (18)$$

Here, $\{v\}^H\{v\}$ is the plate velocity spectrum, $S_{vv}$. The RSP from plate 'b' can be calculated as,

$$RSP(\omega) = [\mathbf{S_{vv,b}}][\mathbf{R}] \qquad (19)$$

*2.4. Transmission loss*

Transmission loss is estimated as the ratio of incident power on the skin panel and the radiating power (RSP) by the trim panel as,

$$TL = 10\log 10\left(\frac{\Phi_{pp}S}{4\rho_{ext}c_{ext}RSP}\right) \qquad (20)$$

**4. Results and discussion**

*4.1. TBL-induced sound radiation: validation*

The developed FE-RRM model is validated by the analytical work reported by Maury et al. [2]. The physical and mechanical properties of the typical aircraft panel and the turbulent flow are considered same as given in Table 1 in p. 1864 of Maury et al. [2]. Two types of panels are used, a) tensioned and b) non-tensioned. FE meshing for flexible panels is adopted as to keep the element size well below the wavelength ($\Delta x < \lambda/3$) of the plate bending wave, in order to account for the convective part of the pressure fluctuations [11]. The in-vacuo free vibration analysis, performed using in-house MATLAB codes, yields the frequency and mode number data, presented in Table 1.

**Table 1** First *eight* frequencies (Hz) and mode numbers

| Tensioned panel | | Non-tensioned panel | |
|---|---|---|---|
| Present FE | Maury et al., 2001 | Present FE | Maury et al., 2001 |
| 269 | 268 | 39 | 40 |
| 353 | 354 | 83 | 83 |
| 469 | 473 | 115 | 115 |
| 500 | 513 | 155 | 154 |
| 558 | 569 | 230 | 224 |
| 610 | 615 | 242 | 235 |
| 645 | 657 | 257 | 252 |
| 759 | 780 | 285 | 277 |

The TBL-induced RSP for the non-tensioned ($\xi = 0.01$) and tensioned ($\xi = 0.01, 0.05$) panels are validated and presented in Fig. 1. All three cases are found to be in very good agreement with the analytical results. [1]

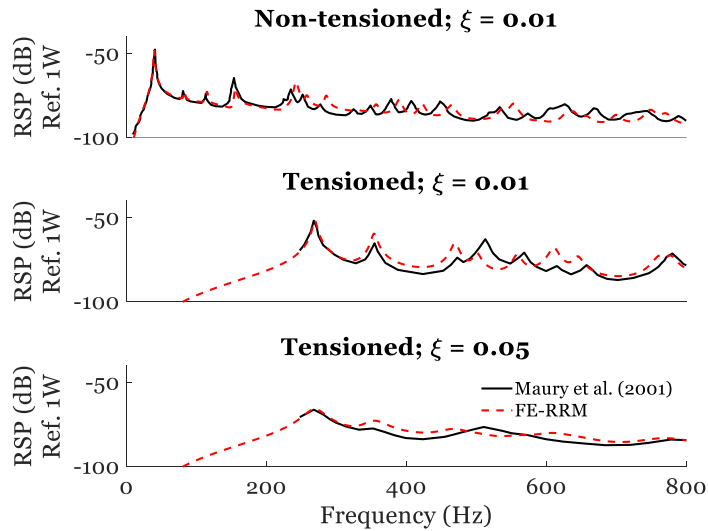

**Fig. 1.** RSP in dB (ref. 1W)

*4.2. Transmission loss through stiffened double-leaf panel*

Once the FE-RRM model is validated, RSP from unstiffened and stiffened double-walled panel system is estimated using the developed mathematical formulation. Both the panels are considered non-tensioned. The mechanical properties of the panels and the physical property of the flow is kept same as in the previous case. Only thickness of the panel a (ta) is varied as 1 mm and 1.5 mm. The cavity depth is 0.1 m. The cavity is discretized such that the first several cavity modes can be captured. Two different stiffener (0.02m deep) configurations, i) along flow at mid-width, ii) along cross-flow at mid-length are used for the excited skin panel only. The TLs in the 1/3$^{rd}$ octave band for different configurations are presented in Fig. 2.

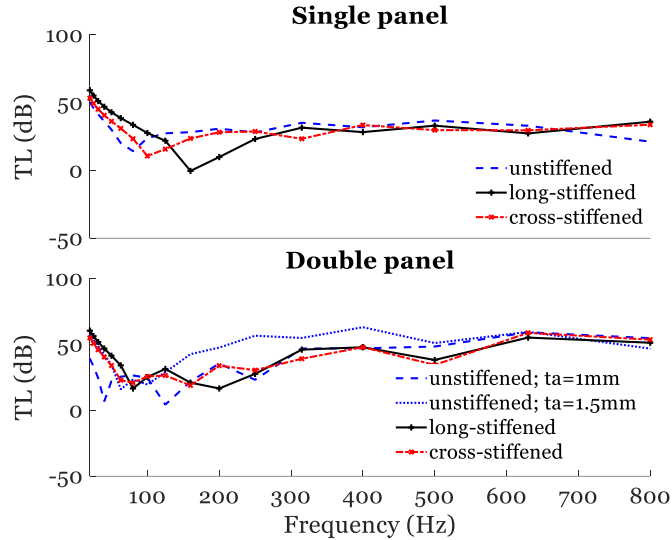

**Fig. 2.** Transmission loss for different panel configurations

It is observed that, as always, the double wall panel systems work better as sound insulator (higher TL values) across all the frequencies. However, it is seen that for the given geometrical configuration studied (1) beyond 250 Hz the sound insulating behavior of the stiffened and the unstiffened panels are almost identical in the 1/3$^{rd}$ octave band, (2) the double wall panel with unequal panel thickness works as a better insulator, (3) beyond 600 Hz all the combinations of the double wall panel systems behaves very similar in terms of TL.

It is worth noting that 70 panel 'a' modes, 80 panel 'b' modes and 50 cavity modes are considered throughout. On vectorization of MATLAB codes, the fully coupled system is solved with 4.5s CPU time per 1Hz frequency in a DELL Workstation with 8 cores and 32GB RAM.

## 5. Conclusion

A coupled numerical model using FE-RRM technique is proposed and successfully implemented to predict energy transmission through TBL-excited stiffened double-leaf panel system. In general, the double panel system works as a better sound insulator than the single panel as expected. There is a scope for exploring the effectiveness of stiffeners by optimizing the location and orientation of the stiffeners without compromising on the total mass of the system. This work can further be extended for stiffened panel-cavity-panel-enclosure problem for enclosure SPL estimation.

**CRediT authorship contribution statement**

**Biplab Ranjan Adhikary:** Conceptualization, simulation, validation, manuscript preparation, review & editing. **Atanu Sahu:** Conceptualization, review and editing. **Partha Bhattacharya:** Conceptualization, review and editing.

**Declaration of Competing Interest**

The authors declare that they have no known competing financial interests or personal relationships that could influence the work reported in this paper.